\documentstyle[aps,prl,epsfig]{revtex}
\begin{document}
\wideabs{
\title{Columnar Defects and Scaling Behavior in Quasi 2D Type II
Superconductors}
\author{Gregory M. Braverman}
\address{Max Planck Institut f\"ur Kernphysik, Heidelberg, Germany}
\author{Sergey A. Gredeskul and Yshai Avishai}
\address{Ben-Gurion University of the Negev, Beer-Sheva, Israel}
\date{\today}
\maketitle
\begin{abstract}
Persistent scaling behavior of magnetization in layered 
high $T_c$ superconductors with 
short--range columnar defects is explained within the 
Ginzburg Landau theory.
In the weak field region, the
scaling function differs from that of a clean sample and both 
the critical and crossing temperatures are 
renormalized due to defects. 
In the strong field region, defects are 
effectively suppressed and 
scaling function, as well as critical 
and crossing temperatures are the same as in a clean 
superconductor.
This picture is consistent with recent experimental results
\end{abstract}
}
Lay\-ered high-tem\-pe\-ra\-tu\-re su\-per\-con\-duc\-ting (H\-T\-S\-C)
materials, such as
Bi$_2$\-Sr$_2$\-CaCu$_2$\-O$_{8+\delta}$ and
Bi$_2$\-Sr$_2$\-Ca$_2$\-Cu$_3$\-O$_{10}$,
are known to exhibit
2D scaling magnetic properties\cite{kes,tesa,li,tesb} around the mean
field transition line $H_{c2}(T)$. It is 
manifested by inspecting the magnetization $M_{0}$ as a
function of temperature $T$ (measured in energy units) and the
(external)
magnetic field
$H$\cite{tesb}:
\begin {equation}
\frac{s\Phi_0}{A\sqrt{TH}}M_0(T,H)=-(\sqrt{x^2+2}-x),
\label{Magn_0}
\end{equation}
where $s$ is an effective interlayer spacing, $\Phi_0$ is
the flux quantum, $x=AH_{c2}^\prime[T-T_{c2}(H)]/\sqrt{TH}$
 is the scaling 
variable, $H_{c2}^{\prime}\equiv-dH_{c2}(T)/dT|_{T=T_{c0}},$  and
$T_{c0}$ 
is the zero field critical temperature. For a superconductor with
Ginzburg-Landau (GL)  
parameter $\kappa$ and Abrikosov geometric factor  $\beta_{A}$ the 
constant $A=\sqrt{s\Phi_0/p}$, where   
$p=16\pi\kappa^2\beta_A.$ 
The form of the scaling
function \cite{tesb} implies the 
existence of a crossing point: at some temperature 
$T^{*}_{0}=T_{c0}(1+1/(2A^{2}H_{c2}^\prime))^{-1},$
the sample magnetization is {\em independent} on $H$,
$M_0^*\equiv M_{0}(T^{*}_{0},H)=-T^{*}_{0}/(s\Phi_0)$.

Recently,
the influence of linear defects (columnar defects, artificial holes
etc.) 
on the magnetic properties of  
superconductors has been studied
experimentally\cite{civale,moshchalkov,beek} and 
theoretically\cite{shapiro,buzdin,tes1,we,we1}.
In particular, experiments by van der Beek {\it et.
al.}\cite{beek} showed that in HTSC with columnar defects
the reversible
magnetization of the sample is 
drastically affected, and that there are now {\it two}
scaling regimes pertaining to relatively weak $H< H_\Phi$ and strong 
$H>H_\Phi$ magnetic fields (here the
matching field $H_\Phi=n_d\Phi_0$ is proportional to the 2D density 
of defects $n_{d}$). These two scaling regimes correspond to two
different critical temperatures (used in Ref.\cite{beek} 
as fitting parameters) and 
crossing points.\\
\phantom{aaa}In this Letter we propose an explanation of these
results. Let us commence by presenting some intuitive arguments.
Consider the quantity
$c=H_{\Phi}/H$ which, in a macroscopic sample, 
is the number of defects 
divided by the number of vortices. 
The magnetic field then serves as a control parameter for tuning
the effective concentration $c$ of defects. In the weak 
field region, $c$ is large,
 each vortex is affected by a force emanating from many defects, and 
 the fluctuations of this force play the main role. Short-range  
 defects could be taken into account perturbatively. In first 
 order they retain the same form of 
  scaling function as that of a clean sample but renormalize the
critical 
 temperature $T_{c}$. Second order corrections indeed
destroy the scaling behavior but in the vicinity 
 of the crossing temperature scaling is approximately 
 maintained. In the strong 
 field region, $c$ is small, and the standard 
 concentration expansion \cite{IM} can be used. Here, even 
 the first order correction (with respect to small concentration) 
 destroys the scaling behavior. However, 
 a strong field effectively suppresses the defects,
 thus restoring the scaling behavior of a clean superconductor 
 with the initial critical temperature $T_{c0}.$ 
 Identifying the two fitting temperatures of 
 Ref.\cite{beek} with the renormalized critical temperature $T_{c}$ 
 and the initial one $T_{c0}$ respectively, one finds for the
 dimensionless defect strength $\theta_{1}=0.49$, well 
 inside its allowed range $0 \le \theta_{1} \le 1$. 
 This indicates a full consistence
 between the description constructed below and the experimental
  results of Ref.\cite{beek}.\\   
\phantom{aaa}Our quantitative 
discussion employs an approach 
proposed and successfully used for arbitrary fields 
in clean superconductors
\cite{tesa,tesb,tesan} and for very 
low fields in disordered 
superconductors\cite{tes1}. Here we use it for 
disordered superconductors in much higher fields.
Consider an irradiated thin
superconducting film (or one layer in a layered
superconductor) with area $S$ subject to perpendicular
magnetic field (thus parallel to the defects). 
The effective interlayer separation 
$s$ is assumed to be much larger than
the effective superconducting coherence length $\xi(H,T)$ 
in the magnetic field direction but much smaller 
than the magnetic penetration depth. Then
the problem becomes effectively two dimensional\cite{tesan}.
Columnar defects can be 
described as a local
reduction of the critical temperature
$\delta T_c({\bf r})=T_{c0}\sum t_j\exp(
-({\bf r}-{\bf r}_j)^2/2L^2).$
Here ${\bf r}$ is a two dimensional
vector in the film plane, $L$ is the 
defect radius, and the positions ${\bf r}_j$ of defects are 
uniformly and
independently distributed over the film plane with density $n_d.$ The 
value of $n_d$ is assumed to be moderate so that for 
the pertinent region of temperature the matching 
field $H_{\Phi}$ is always much smaller than $H_{c2}(T).$  
The dimensionless amplitudes of defects $t_j\leq 1$
are also independent random quantities distributed with 
some probability density.
The thermodynamic properties of a  
type-II superconductor with $\kappa\gg 1$
containing $N_{v}$ vortices 
are described by its partition function
\begin{equation}
{\cal Z} \propto 
\int{{\cal D}\{\Psi\}
\exp(-N_vg[\Psi])},
\label{PartFunc}
\end{equation}
where $\Psi$ is the corresponding order parameter. 
The dimensionless GL free energy $g[\Psi]$ of an irradiated
superconductor 
is given by an expression
\begin{equation}
g=
x\overline{|\Psi|^2}+(4\beta_A)^{-1}\overline{|\Psi|^4}
+\overline{\tau|\Psi|^2},
\label{FrEn}
\end{equation}
where bar denotes averaging over the sample area. 
The scaling variable $x$ and
the local temperature $\tau({\bf r})$ are defined below
for each region of the magnetic field.\\
\phantom{aaa}Following\cite{tes1}, we replace 
in Eq.(\ref{FrEn}) $\overline{\left|\Psi({\bf
r})\right|^4}$ by $\beta_A\left(\overline{\left|\Psi({\bf
r})\right|^2}\right)^2,$ where $\beta_{A}\sim 1.16$ 
is the Abrikosov factor for a
triangular lattice. This  
replacement is based on the assumption that the distribution of 
vortices is almost uniform in both regions of the magnetic field 
considered here. It is
supported by noticing a remarkable difference between the 
number of vortices and the number of defects in both regions of 
fields\cite{Wwe}.
This substitution,
together with the simplest version of the
Hubbard-Stratonovich transformation 
(introduction of an additional integration 
over some auxiliary field $\gamma$) turns 
the problem to be an exactly 
solvable one\cite{tes1}. Then project the order 
parameter on the lowest Landau level (LLL) subspace, 
\begin{equation}
\Psi({\bf r})={\displaystyle \sum_{m=0}^{N_{v}}} C_m L_m ({\bf r}),
\label{Eq_LLL}
\end{equation}
where $L_m ({\bf r})$ are 
normalized LLL eigenfunctions with orbital momentum $m.$
As  was recently demonstrated\cite{liRos}, the LLL approximation works 
quite well even down to $H \ge H_{c2}(T)/13$.
After integration over the expansion coefficients $C_{m}$ the partition 
function (\ref{PartFunc}) reads,
\begin{equation}
{\cal Z}\propto\int_\Gamma
\exp
\left\{-N_v{\cal L}(\gamma,x)
\right\}d\gamma,
\label{PartFuncGamma}
\end{equation}
where
\begin{equation}
{\cal L}(\gamma,x)=-\gamma^2+N_v^{-1}\textrm{tr}\ln
\left[
(x+\gamma)\hat{\textrm{I}}+\hat{\tau}
\right]
\label{action}
\end{equation}
and $\hat{\tau}$ is a random matrix with 
elements:
\begin{equation}
\tau_{mn}=\int_S L_m^*({\bf r})\tau({\bf r})
L_n({\bf r})d^2{\bf r}.
\label{TauMN}
\end{equation}
The contour $\Gamma$ in Eq.(\ref{PartFuncGamma}) 
is parallel to the imaginary axis and
stretches from $\gamma^*-i\infty$ to  $\gamma^{*}+i\infty.$
To assure convergence of the integrals over the
coefficients $\{C_m\}$ the real
constant $\gamma^{*}$ should satisfy the inequality $\gamma^*+x+\min 
\tau_{n}>0,$ where $\tau_{n}$ is the $n$-th eigenvalue of the matrix 
$\tau_{mn}$.\\
\phantom{aaa}In the thermodynamic limit 
$S\rightarrow\infty$ with  $n_{d}=N_{v}/S$ fixed, the partition 
function (\ref{PartFuncGamma}) could be calculated in a saddle 
point approximation. This results in the following form for 
the magnetization
\begin{equation}
\frac{s\Phi_0}{A\sqrt{HT}}M(T,H)=
\left(N_{v}^À{\cal Z}\right)^{-1}\partial {\cal Z}/\partial x=
-2\gamma(x),
\label{MagnGen}
\end{equation}
where $\gamma(x)$ is the 
solution of the saddle point equation 
$\partial{\cal L}(\gamma,x)/\partial\gamma=0$.  
For a clean superconductor 
($\hat{\tau}=0$)  one gets two possible saddle
points but only one of them 
\begin{equation}
	\gamma_0 (x)=\frac{1}{2}(\sqrt{x^2+2}-x)
	\label{gamma_0}
\end{equation}
can be reached by an allowed deformation of the
contour $\Gamma$. Substitution of Eq.(\ref{gamma_0}) into 
(\ref{MagnGen}) yields the magnetization $M_{0}(T,H)$ of a clean sample 
(\ref{Magn_0}) obtained in Ref.\cite{tesb}. Note that 
$-2\gamma_{0}(x)$ serves as the appropriate scaling function.
To study the disordered case, we consider separately two 
 regions of the magnetic field.\\
\phantom{aaa}In the weak field region $H< H_\Phi$ we,
from the onset, take into 
account the renormalization of the critical temperature caused by
defects.
As a result, the scaling variable $x$ is defined in the 
same way as for a clean superconductor albeit with renormalized critical 
temperature  
$T_{c}=T_{c0}-\delta T_{c},$ where  
\begin{equation}
	\delta T_{c}=<\delta T_{c}({\bf r})>=2\pi\theta_{1} n_{d}L^{2}T_{c0}
	\label{deltaT}
\end{equation}
and $\theta_{n} \equiv <t^{n}>$ where here and below 
$<..>$ implies ensemble average.
The function $\tau({\bf r})$ in this field region, defined as 
$\tau({\bf r})=(\delta T_{c}({\bf r})-\delta 
T_{c})AH_{c2}^{\prime}/\sqrt{TH}$, represents 
temperature fluctuations 
caused by short-range defects. They are
small and can be accounted for perturbatively. 
Then, in the thermodynamic limit, 
the last term on the r.h.s. of Eq.(\ref{action})
has an explicit self-averaged structure
$N_v^{-1}\textrm{tr}(...)$ and can be replaced by its
average. This procedure modifies the saddle point equation
and therefore  results in a modified magnetization
\begin{equation}
M(T,H)=M_0(T,H)\left(
1+\varepsilon (T)\frac{2\gamma_0(x)}{\sqrt{x^2+2}},
\right),
\label{MagnWeakNew}
\end{equation}
where
\begin{equation}
	\varepsilon (T)=\left<\frac{{\rm 
	tr}\hat{\tau}^{2}}{N_{v}}
	\right>=
	\frac{\theta_{2}}{p}n_dL^2
	\frac{(2\pi H_{c2}^{\prime}T_{c0})^{2}sL^{2}}{T}.
    \label{varepsilon}
\end{equation}
Note that the parameter $\varepsilon(T)$ is proportional to the fourth 
power of the defect radius $L$ thus justifying the 
perturbation approach for short-range defects.\\
\phantom{aaa}In the zeroth approximation with respect to $\varepsilon
(T)$ 
the magnetization (\ref{MagnWeakNew}) has exactly the same
form, as for a clean sample (\ref{Magn_0}) thus retaining
both the scaling property and 
the existence of a crossing point. However, due to renormalization 
of the critical temperature, the crossing temperature 
$T^{*}=T_{0}^{*}-\delta T^*$ differs from its 
value $T_{0}^{*}$ in a clean sample:
$\delta T^* =
\delta T_{c}(1+(2A^{2}H_{c2}^\prime)^{-1})^{-1}.$
In the next order, scaling is virtually destroyed, since the
correction term (within the parenthesis in Eq.(\ref{MagnWeakNew}))
depends not only on the scaling variable $x$ 
but also on temperature. Yet, in a sufficiently narrow 
region around some temperature $T$, 
 the deviation from scaling is
negligibly small, but the scaling function itself is modified to be
$-2\gamma(x,T)$. At temperature $T^{*}$ the magnetization reads
\begin{equation}
	M(T^*,H)=M_0(T^{*})\left(1+\varepsilon (T^*)\frac{2H^*}
	{H+H^*}\right),
	\label{Magn_3}
\end{equation}
where $H^{*}=H_{c2}(T^{*})=T^{*}/(2A^{2}).$
Therefore if the field is weak enough, $H\ll H^{*},$ then the crossing 
point is 
restored, $T^{*}$ serves as a true crossing temperature and 
the magnetization at the crossing temperature  
differs from its unperturbed form $-2\gamma_{0}(x)$ merely by a 
multiplicative constant $1+2\varepsilon(T^{*}).$\\
\phantom{aaa}When the magnetic field 
increases, the approach used above becomes 
inapplicable. Firstly,
it fails in the vicinity of the matching field where the Abrikosov 
factor becomes very sensitive to the details of  
defect configuration. Secondly,
 higher order terms in the perturbation expansion for the 
saddle point equation (which are omitted), grow with magnetic 
field. Fortunately, we have here a new 
small parameter, that is, the
dimensionless concentration $c$ of defects. It is then
natural to use the concentration expansion. In such a case there
 is no sense in renormalizing the critical temperature, and the
 dimensionless temperature $\tau({\bf r})$ is now defined as
$\tau({\bf r})=\delta T_{c}({\bf r}) AH_{c2}^{\prime}/\sqrt{TH}.$\\
\phantom{aaa}As mentioned above, 
the second term in the r.h.s. of Eq.(\ref{action}) is self-averaging 
and can be calculated using the 
limiting form of the density of states $\rho(\tau)$ of the 
matrix (\ref{TauMN}), which,
for short-range defects in linear approximation 
 with respect to $c$, reads 
 \begin{equation}
 		\rho(\tau)=(1-c)\delta(\tau)+
\frac{c}{\lambda}p(\frac{\tau}{\lambda}),
 	\label{DOS}
 \end{equation}
 where $\lambda=
 2\pi L^{2}T_{c0}A H_{c2}^{\prime}\sqrt{H}
 (\Phi_{0}\sqrt{T})^{-1}$ and $p(t)$ is probability distribution of the
 dimensionless
 temperature $t_j$. Indeed, the matrix $\tau_{mn}$ is nothing but  
 the Hamiltonian of a particle with charge $2e$ in a $2D$ system 
 subject to a perpendicular magnetic field and containing  short-range 
 defects (projected on the LLL). 
 The first and second terms in Eq.(\ref{DOS}) 
 correspond, respectively, to  
 those states whose energy is stuck to the LLL (despite the presence of 
 zero-range defects (see {\em e.g.}\cite{GZAA})) and those 
  states whose energies are 
  lifted from the LLL by these defects. 
  For sufficiently 
  narrow 
  distribution $p(t)$, the corresponding saddle-point 
  equation leads
to the magnetization
\begin{equation}
	M=M_{0}\left(
	1-\frac{c\lambda\theta_{1}}{(1+2\lambda\theta_{1}\gamma_{0}(x))\sqrt{x^
{2}+2}}
	\right),
	\label{Magn_4}
\end{equation}
were $M_{0}(T,H)$ is given by Eq.(\ref{Magn_0}) with an initial 
critical temperature $T_{c0}.$\\
\phantom{aaa}Rigorously speaking, scaling is destroyed since 
both the concentration 
$c$ and the shifted eigenvalue $\theta_1\lambda$ depend explicitly on 
$H$ and $T$. However, at strong field the 
correction term in Eq.(\ref{Magn_4}) becomes negligibly small. This 
implies a restoration of the crossing point. Indeed, at
temperature $T_{0}^*$ 
the magnetization
$M^*=M(T^*,H)$ assumes the form
\begin{equation}
	M^*=M_0^*
\left(1-
\frac{1}{1+\eta}\frac{H_\Phi}{H+H^{*}}
\right),
	\label{Magn_5}
\end{equation}
with $\eta^{-1}=2\pi
L^2H_{c2}^\prime T_{c0}\theta_1/\Phi_0.$ Therefore in the entire strong 
field region $H_{\Phi}\ll H\ll H^{*}$ the crossing temperature coincides 
with its initial value $T_{0}^{*}$ and the magnetization in the crossing 
point  practically coincides with its  
value $M_{0}^{*}$ in a clean superconductor.\\
\phantom{aaa}Let us now discuss the 
limits of applicability of our results and 
their relation to the experiment of Ref.\cite{beek}. Note that the first 
two moments $\theta_{1,2}$ of the random dimensionless temperature $t$ 
satisfy the inequality $0\leq\theta_{1}^{2}\leq \theta_{2}\leq 1.$ 
In the pertinent region of fields\cite{beek} 
$0.2\div5 {\rm T}$, a typical defect radius 
 $L \sim 3.5{\rm nm}$ is at least one order of magnitude smaller than 
 the magnetic length, hence the defects can definitely 
 be taken as short-range ones. 
In the weak field region, the  
important small parameters are then $\varepsilon(T)$ (which enters the 
magnetization (\ref{Magn_3})) and $\varepsilon (T)/(x+\gamma_{0}(x))^
{2}$ 
(which enters the saddle point equation).  
Using parameters from the experimental setup\cite{beek} 
$s=1.5{\rm nm}$, $k_{B}\mu_0 H_{c2}^\prime=1.15{\rm TK}^{-1}$, $\kappa=
100$,
$n_d=5\times10^{10}{\rm cm}^{-2}$, $\mu_0H_\Phi=1{\rm T},$ 
$T=75\div 85 {\rm K},$ $T^{*}=78.9{\rm K},$ we find from 
Eq.(\ref{varepsilon}) 
$\varepsilon(T^{*})=0.5\theta_{2}$ 
and $\varepsilon (T^{*})/(x^{*}+\gamma_{0}(x^{*}))^{2}\approx 0.25$ 
(the latter figure is obtained for $\mu_{0}H=0.2{\rm T}$). For quite 
plausible value $\theta_{2}=0.5$ one then finds
$\varepsilon(T^{*})=0.5\theta_{2}=0.25.$ The condition of convergence 
of the integral over the expansion coefficients $\{C_{m}\}$ can be
written 
as $H>0.25H_{\Phi}\theta_{1}^{2}/\theta_{2}$ and even in the worst 
case $\theta_{1}^{2}=\theta_{2}$ it reads $\mu_{0}H\approx 0.25{\rm T}.$ 
Finally, one has $\mu_{0}H^{*}\approx 6.4 {\rm T}$ and applicability of 
the LLL projection requires $\mu_{0}H>0.5 {\rm T}.$ The weak 
field region of Ref.\cite{beek} corresponds to 
$\mu_{0}H=0.2\div0.02 {\rm T}.$ Thus, in the weak 
field region, the condition for
applicability of the LLL projection is
slightly violated, but the deviation
is not dramatic. In the strong field region we find 
$\eta\approx 2.9$ and therefore the correction term in parenthesis of
equation (\ref{Magn_5}) is less than three percents. Hence, in this
region 
our assumptions are fully satisfied.\\
\phantom{aaa}Using the same set of parameters we display 
in figure 1 the quantity $M/\sqrt{TH}$ as a 
function of the scaling variable for weak field (inset) 
and strong field (main part). 
We used here the maximal value $\theta_{1}=1$. 
In the strong field region, the 
deviation form clean sample scaling behavior is negligibly small 
for all three values of strong magnetic field.
\begin{figure}[t]
\begin{center}
\epsfxsize=8cm 
\epsfbox{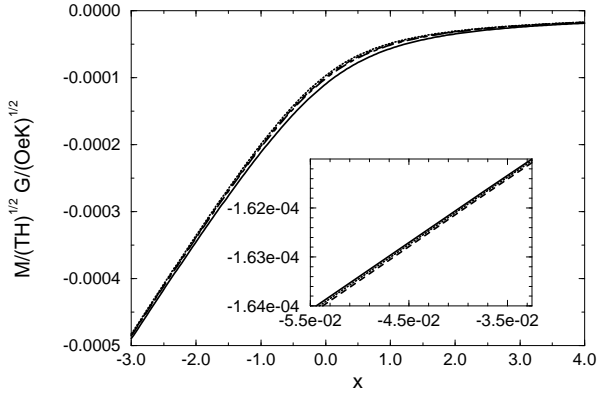}
\end{center}
\caption{The quantity $M/\sqrt{HT}$, as a function of $x$. 
Dashed, dotted and dot-dashed lines correspond to 
$\mu_0 H=$ $3T$, $4T$ and $5T$ (main figure) and correspond to $\mu_0 H=
$ $0.02T$, 
$0.1$ and $0.2T$ (insertion). The solid line corresponds to the
clean--sample scaling function (strong field region only).}
\label{fig1}
\end{figure}
In the weak field region, the scaling
functions for three different fields can hardly be distinguished. 
This means that scaling is undoubtedly valid in a 
vicinity of the crossing temperature. 
At the same time the scaling function 
differs from its form in a clean sample (\ref{Magn_0}) by a
multiplicative
constant (see the parenthesis in Eq.(\ref{MagnWeakNew})). Note that 
scaling in the weak field region 
(which was experimentally established) is
less pronounced than that in the strong field region. 
Apparently, the reason 
is that the experimental data are fitted to 
account for the clean sample scaling 
function. Nevertheless if we identify the  fitted temperature 
$82.6 {\rm K}$ (found in Ref.\cite{beek} 
in the weak field region) with 
the renormalized critical temperature 
$T_{c}=T_{c0}-\delta T_{c}$, and the 
fitted critical temperature 
$84.2{\rm K}$ in the strong 
field region with $T_{c0}$, then, even within such a rough
approximation, 
we obtain $\theta_{1}\approx 0.5.$ Recalling that 
$\theta_{1}$ should be positive 
and less than unity, the 
above result strongly supports the
applicability of our theory to the pertinent experiment\cite{beek}.\\
\phantom{aaa}In summary, we
calculated the magnetization of an irradiated
superconductor below the mean--field transition
line $H_{c2}(T)$, using the approach developed 
in Refs.\cite{tesa,tesb,tes1,tesan}. 
It was shown that, from a rigorous point of view,
disordered short-range defects 
are expected to destroy the scaling behavior and prevent the 
existence of crossing point in both 
regions of weak and strong  
magnetic fields (with respect to matching field $H_{\Phi}$). 
And yet, in the framework of the experimental
setup\cite{beek} the deviation from scaling behavior appears to be
negligibly small and crossing points exist in both field regions, in 
complete agreement with the experimental 
findings. The two fitting critical 
temperatures introduced in Ref.\cite{beek} for the 
strong and weak field 
regions correspond, in our formalism,
to the initial and renormalized critical 
temperatures.\\ 
\phantom{aaa}The authors would like to thank 
Z. Tesanovi\'c for helpful discussions and P.H. Kes who 
advised us some parameters of experimental setup of Ref.\cite{beek}.\\
\phantom{aaa}This work was supported by MINERVA Foundation 
(G.B.), by grants from Israel Academy of Science ``Mesoscopic
effects in type II superconductors with short-range pinning
inhomogeneities'' (S.G.), and ``Center of Excellence'' (Y.A.), and by
DIP
grant for German-Israel collaboration (Y.A.).\\ 
 

\begin{thebibliography}{99}
\bibitem{Wwe} However field could not be very strong: if the order
parameter 
is close to zero, vortices could collapse onto a 
single defect\cite{we}. But in the case 
$H_{\Phi}\ll H_{c2}(T)$, there exists a finite region 
of strong fields $H_{\Phi}<H<H_{c2}(T)$ where the 
distribution of the vortices does not feel the presence of defects.
\end{thebibliography}
\end{document}